\documentclass[final]{svjour2}
\usepackage{graphicx}
\usepackage{rotating}
\usepackage{amssymb}
\usepackage{mathptmx}
\usepackage{amsmath}
\usepackage[numbers]{natbib}
\usepackage[FIGTOPCAP]{subfigure}

\makeatletter
\journalname{Journal of Low Temperature Physics}
\def\lesssim{\ \raise.3ex\hbox{$<$}\kern-0.8em\lower.7ex\hbox{$\sim$}\ }
\def\gesim{\ \raise.3ex\hbox{$>$}\kern-0.8em\lower.7ex\hbox{$\sim$}\ }

\bibpunct{}{}{,}{s}{}{,}
\begin{document}

\newcommand{\hdblarrow}{H\makebox[0.9ex][l]{$\downdownarrows$}-}
\title{Superfluid theory of a gas of polarized dipolar Fermi molecules}
\author{Y. Endo$^1$ \and D. Inotani$^2$ \and Y. Ohashi$^1$}
\institute{1:Department of Physics, Faculty of Science and Technology, Keio University, 3-14-1, Hiyoshi, Kohoku-ku, Yokohama 223-8522, Japan\\
%Tel.: +81-45-566-1454 \\ Fax: +81-45-566-1672 \\
%\email{info@yuki-phys.com}
\\2: Graduate School of Pure and Applied Sciences, University of Tsukuba, Tsukuba, Ibaraki 305-8571, Japan}

\date{7,7,2013}
\maketitle
\bibliographystyle{prsty}

\begin{abstract}
We present a superfluid theory of a polarized dipolar Fermi gas. For two dipolar molecules each of which consists of two atoms with positive charge and negative charge, we derive an effective dipole-dipole pairing interaction. Using this pairing interaction, we show that the resulting BCS gap equation is not suffered from the well-known ultraviolet divergence, so that one can quantitatively predict superfluid properties of a dipolar Fermi gas. Using this cutoff-free superfluid theory, we examine the symmetry of the superfluid order parameter at $T=0$. We also discuss the deformation of the Fermi surface, originating from the anisotropy of the dipole-dipole interaction.

\end{abstract}
%%%%%%%%%%%%%%%%%%%%%%%%%%%%%%%%%%%%%%%%%%%%%%%%%%%%%%%%%%%%%%%%%%%%%%%%%%%%%%

\section{Introduction}
Recently, ultracold gases of Fermi molecules have attracted much attention. This system has a long range dipole-dipole interaction associated with a molecular electric dipole moment. In addition, this interaction is anisotropic. Thus, one expects richer physics than the case of the simplest isotropic $s$-wave contact interaction, which has been dominantly examined in cold Fermi gas physics. In particular, it has been predicted that the dipole-dipole interaction induces an unconventional $p$-wave superfluid~\cite{Baranov2002_2,Baranov2004,Baranov2008,Bruun2008,Zhao2010,Shi2010,Levinsen2011}. Although the $p$-wave superfluid has already been realized in liquid $^3$He, what is expected in a dipolar Fermi gas is the polar phase~\cite{Baranov2002_2,Baranov2008}, which has not been realized even in liquid $^3$He. Although this exciting prediction has not been experimentally confirmed, some groups have succeeded in creating heteronuclear molecules by using a Feshbach resonance~\cite{Wu2012,Heo2012}, that are stable against the chemical reaction~\cite{Ni2010}. Thus, although various difficulties, such as the cooling problem, still remain, the superfluid phase transition of a dipolar Fermi gas might be achieved near future.
\par
Since the superfluid phase transition of a dipolar Fermi gas has not been realized yet, quantitative prediction of the superfluid phase transition temperature $T_{\rm c}$ is an important issue in the current stage of theoretical research. In this regard, we note that, when we employ the BCS theory for a dipolar Fermi gas, the BCS gap equation is known to exhibit ultraviolet divergence~\cite{Baranov2002_2,Baranov2008,Zhao2010,Shi2010}. In the case of the conventional $s$-wave interaction, this divergence can be eliminated by introducing the $s$-wave scattering length~\cite{Randeria1995}. However, such a renormalization is difficult in the dipolar case, because the superfluid order parameter actually consists of various symmetry components belonging to the odd parity~\cite{Baranov2002_2,Baranov2008,Zhao2010,Shi2010,Levinsen2011}. In addition, since an electric dipole-dipole interaction in the present system is usually strong~\cite{Wu2012,Heo2012,Ni2008,Deiglmayr2008}, a strong-coupling treatment is necessary beyond the simple mean-field level, including the momentum dependence of the dipole-dipole interaction. Thus, although several groups have theoretically discussed the superfluid instability of this system~\cite{Baranov2002_2,Baranov2008,Zhao2010,Shi2010}, the quantitative evaluation of $T_{\rm c}$ for arbitrary strength of the dipole-dipole interaction still remains as an important challenge.
\par
Toward to quantitative analysis of the superfluid phase transition in a dipolar Fermi gas, in this paper, we present a superfluid theory for a polarized dipolar Fermi gas which does {\it not} involve the ultraviolet divergence from the beginning. Because of this advantage, this theory does not need a high-energy cutoff, nor the renormalization, so that we can conveniently discuss various superfluid physics only using physical parameters involved in the starting Hamiltonian. To see how this theory works, we examine the superfluid state at $T=0$ within the framework of the BCS-Leggett theory~\cite{Leggett1960}. In this paper, we take $\hbar=k_{\rm B}=1$, and the system volume $V$ is taken as unity, for simplicity.
\par
%============================================================================
\begin{figure}[t]
\begin{center}\sidecaption
{\includegraphics[width=0.35\linewidth,keepaspectratio]{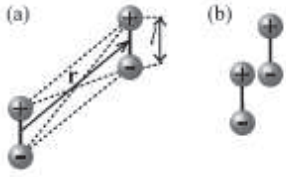}}
\caption{(a) Model dipole-dipole interaction between polarized dipolar molecules. Each molecule consists of a fermion with positive charge and a boson with negative charge. $l$ is a molecular size. $\textbf{r}$ points from the center of the left molecule to that of the right molecules. The dipole-dipole interaction in Eq.~(\ref{bare_dipolar_r}) is obtained by the sum of four Coulomb interactions (dashed lines). (b) Effective interaction between dipolar molecules coming from a contact interaction $g_{\rm BF}$ between Fermi and Bose atoms.}
\label{fig:DipoleDipole}
  \end{center}
\end{figure}
%============================================================================

%--------------------------------------------------------------
\section{Model dipole-dipole interaction between Fermi molecules}
We consider an interaction between two model heteronuclear molecules, each of which consists of a Fermi atom charging positively ($+Q$) and a Bose atom charging negatively ($-Q$) with the distance $l$, as schematically shown in Fig.\ref{fig:DipoleDipole}(a). They are assumed to be polarized in the $z$-direction by an external electric field $\textbf{E}=(0,0,E)$. The electric dipole-dipole interaction between the molecules is given by the sum of four Coulomb interactions shown in Fig.\ref{fig:DipoleDipole}(a), which gives
\begin{eqnarray}
U\left(\textbf{r}\right)=\frac{Q^2}{4\pi\epsilon_0}\left[\frac{2}{r}-\frac{1}{\sqrt{r_\perp^2+\left(z-l\right)^2}}-\frac{1}{\sqrt{r_\perp^2+\left(z+l\right)^2}}\right],
\label{bare_dipolar_r}
\end{eqnarray}
where the vector $\textbf{r}=(r_\perp,\theta_r,z)$ points from the center of the left molecule to that of the right molecule in the cylindrical coordinate. (See Fig.\ref{fig:DipoleDipole}(a).) $\epsilon_0$ is the vuccum permittivity. In momentum space, one finds,
\begin{eqnarray}
U(\textbf{q})=\int d \textbf{r} \ U(\textbf{r}) e^{-i\textbf{q}\cdot\textbf{r}}
={2Q^2 \over \epsilon_0 q^2}
\Bigl[1-\cos(ql\cos\theta_\textbf{q})\Bigr],
\label{bare_dipolar_p}
\end{eqnarray}
where $\theta_{\textbf{q}}$ is the angle of $\textbf{q}$, measured from the $q_z$ axis~\cite{Chan2010}. We briefly note that Eq.~(\ref{bare_dipolar_p}) is positive definite.
\par
Equation (\ref{bare_dipolar_p}) is quite different from the ordinary expression used in this field \cite{Goral2000},
\begin{eqnarray}
V_{\rm dd}(\textbf{q})=\frac{\left(Ql\right)^2}{3\epsilon_0}\left(3\cos^2\theta_{\textbf{q}}-1\right),
\label{approximate_dipolar_p}
\end{eqnarray}
which takes both positive and negative values, only depending on the direction of $\textbf{q}$. This popular expression is obtained from the Fourier transformation of the approximated expression of Eq.~(\ref{bare_dipolar_r}) in the limit $l\ll r$,
\begin{eqnarray}
U(\textbf{r}){\rightarrow}\frac{\left(Ql\right)^2}{4\pi\epsilon_0 r^3}\left[1-3\frac{z^2}{r^2}\right].
\label{approximated_dipolar_r}
\end{eqnarray}
Although the assumption $l\ll r$ looks reasonable for a real dipolar molecule, this condition is actually not always satisfied in the Fourier transformation to obtain Eq.~(\ref{bare_dipolar_p}), because it involves the spatial integration over the region $l\gg r$, where the approximated expression in Eq.~(\ref{approximated_dipolar_r}) is {\it not} correct.
\par
We note that Eq.~(\ref{approximate_dipolar_p}) only depends on $\theta_\textbf{q}$, so that, when it is used in the BCS gap equation, we meet the ultraviolet divergence. In contrast, since $U(\textbf{q})$ in Eq.~(\ref{bare_dipolar_p}) has the factor $q^{-2}$, the resulting BCS gap equation is not suffered from this divergence at all, without introducing a cutoff parameter.
\par
When we include a contact interaction $g_{\rm BF}$ between the Bose and Fermi atoms, it also causes another effective interaction $U_{\rm BF}(\textbf{q})$ between dipolar molecules. (See Fig.~\ref{fig:DipoleDipole}(b).) In momentum space, one has
\begin{eqnarray}
U_{\rm BF}\left(\textbf{q}\right)=2g_{\rm BF}\cos\left(q l\cos\theta_p\right).
\label{eq.BF}
\end{eqnarray}
We briefly note that a contact interaction $g_{\rm BB}$ between Bose atoms does not work in the present polarized case, because, when two bosons meet at the same place, two Fermi atoms also come to the same spatial position, which is, however, prohibited by the Pauli's exclusion principle.
\par

%--------------------------------------------------------------
\section{Cutoff-free BCS-Leggett theory for a polarized dipolar Fermi gas}
\par
To see how the interaction in Eq.~(\ref{bare_dipolar_p}) works, we consider the superfluid phase of a polarized Fermi gas within the framework of the BCS-Leggett theory~\cite{Leggett1960}. For simplicity, we ignore the interaction in Eq.~(\ref{eq.BF}) in what follows. The model Hamiltonian is given by 
\begin{eqnarray}
H=\sum_{\textbf{p}}\epsilon_{\textbf{p}}a_{\textbf{p}}^\dagger a_{\textbf{p}}+\frac{1}{2}\sum_{\textbf{k},\textbf{p},\textbf{q}}U\left(\textbf{q}\right)a_{\textbf{k}+\textbf{q}}^\dagger a_{\textbf{p}-\textbf{q}}^\dagger a_{\textbf{p}}a_{\textbf{k}},
\label{eq.1}
\end{eqnarray}
where, $a_{\textbf{p}}$ is an annihilation operator of a dipolar fermion with kinetic energy $\epsilon_{\textbf{p}}\equiv\textbf{p}^2/{2m}-\mu$, measured from the chemical potential $\mu$, $U\left(\textbf{q}\right)$ is the dipole-dipole interaction in Eq.~(\ref{bare_dipolar_p}). In the mean-field theory, Eq.~(\ref{eq.1}) is reduced to (ignoring unimportant constant terms),
\begin{eqnarray}
H_{\rm MF}=\sum_{\textbf{p}}\xi_{\textbf{p}}a_{\textbf{p}}^\dagger a_{\textbf{p}}+\frac{1}{2}\sum_{\textbf{p}}\left[\Delta^*\left({\textbf{p}}\right)a_{-\textbf{p}}a_{\textbf{p}}+\Delta\left(\textbf{p}\right) a_{\textbf{p}}^\dagger a_{-\textbf{p}}^\dagger\right].
\label{MFHamiltonian3}
\end{eqnarray}
Here, $\Delta({\textbf{p}})=\sum_{\textbf{k}}U(\textbf{p}-\textbf{k})\langle a_{-\textbf{k}}a_{\textbf{k}}\rangle$ is the superfluid order parameter, and the kinetic energy $\xi_{\textbf{p}}=\epsilon_{\textbf{p}}+\Sigma\left(\textbf{p}\right)$ involves the mean-field correction, 
\begin{eqnarray}
\Sigma\left(\textbf{p}\right)=-\sum_{\textbf{k}}U\left(\textbf{p}-\textbf{k}\right)\langle a_{\textbf{k}}^\dagger a_{\textbf{k}}\rangle
=-\sum_{\textbf{k}}{U\left(\textbf{p}-\textbf{k}\right)}\frac{1}{2}\left[ {1}-\frac{\xi_{\textbf{k}}}{E_{\textbf{k}}}\right],
\label{self}
\end{eqnarray}
which comes from the Fock term. (Note that the Hartree vanishes identically in the present case.) In Eq.~(\ref{self}), $E_{\textbf{k}}=\sqrt{\xi_{\textbf{k}}^2+\left|\Delta\left(\textbf{k}\right) \right|^2}$ is the single-particle Bogoliubov excitation spectrum. 
In the ordinary BCS-Leggett theory at $T=0$, we solve the BCS gap equation,
\begin{equation}
\Delta\left(\textbf{p}\right)=-\sum_{\textbf{k}}U\left(\textbf{p}-\textbf{k}\right)\frac{\Delta\left(\textbf{k}\right)}{2E_{\textbf{k}}},
\label{Delta_Eq}
\end{equation}
together with the equation for the number $N$ of dipolar molecules,
\begin{eqnarray}
N={1 \over 2}
\sum_{\textbf{k}}
\left[1-\frac{\xi_{\textbf{k}}}{E_{\textbf{k}}}\right].
\label{N_Eq}
\end{eqnarray}
In addition to these, we also solve Eq.~(\ref{self}), to self-consistently determine $\Delta(\textbf{p})$, $\mu$, and $\Sigma(\textbf{p})$. In this procedure, the strength of $U(\textbf{q})$ is specified by the scaled dipolar strength $C_{\rm dd}\equiv m(Ql)^2N^{1/3}/4\pi\epsilon_0$~\cite{Sogo2009} and the dipolar size $l$, that are both observable quantities. We emphasize that, because of the factor $q^{-2}$ in the interaction in Eq.~(\ref{bare_dipolar_p}), any cutoff parameter is not necessary in our theory.
%============================================================================
\begin{figure}[h]
\begin{center}
\includegraphics[width=0.8\linewidth,keepaspectratio]{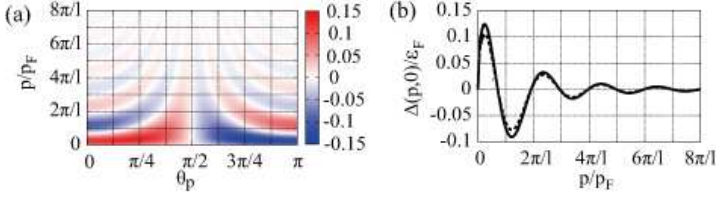}
\caption{(Color online) (a) Calculated superfluid order parameter $\Delta(\textbf{p})=\Delta(p,\theta_\textbf{p})$, normalized by the Fermi energy $\epsilon_{\rm F}$ of a free Fermi gas. $p_{\rm F}=\left(2m\epsilon_{\rm F}\right)^{1/2}$ is the Fermi momentum. (b) $\Delta(\textbf{p})$ as a function of $p$ in the $p_z$ direction ($\theta_\textbf{p}=0$). Dashed line shows the result when $\Sigma\left(\textbf{p}\right)$ in Eq.~(\ref{self}) is ignored. In this figure, as well as in the following figures, we set $C_{\rm dd}=0.22$ and $lp_{\rm F}=0.2$.}
\label{fig:SF}
\end{center}
\end{figure}
%============================================================================
\begin{figure}[t]
\begin{center}
\includegraphics[width=1.0\linewidth,keepaspectratio]{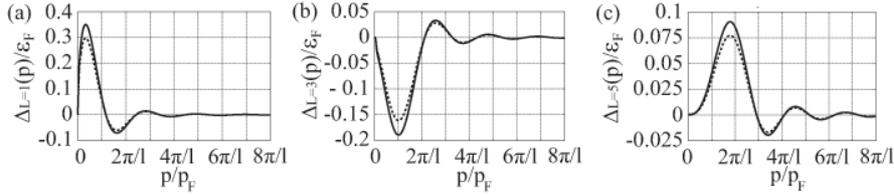}
\caption{Partial wave expansion of $\Delta(\textbf{p})$ in Eq.~(\ref{Delta_L0}). (a) $p$-wave: $\Delta_{L=1}$. (b) $f$-wave: $\Delta_{L=3}$. (c) $h$-wave: $\Delta_{L=5}$. In these figures, dashed lines are the results when $\Sigma(\textbf{p})$ is ignored.}
\label{fig:SF_L0}
\end{center}
\end{figure}
%============================================================================

%--------------------------------------------------------------
\section{Superfluid dipolar Fermi gas at $T=0$}
\par
Figure~\ref{fig:SF} shows the anisotropy of the superfluid order parameter $\Delta(\textbf{p})$. In this calculation, we assume the axisymmetric state with respect to the $p_z$ axis, which has been discussed as a candidate for the pairing state in a dipolar Fermi gas~\cite{Baranov2008,Zhao2010}. Panel (a) shows that $\Delta(\textbf{p})$ is antisymmetric with respect to the $p_x$-$p_y$-plane ($\theta_{\rm p}=\pi/2$). In addition, as shown in panel (b), $\Delta(\textbf{p})$ oscillates in the radial direction $p$, with the period $\simeq 2\pi/l$. As a result, while a line node only exists in the $p_x$-$p_y$-plane when $p\simeq 0$, additional two line nodes appears when $p/p_{\rm F}\gesim \pi/l$. (See Fig.~\ref{fig:SF}(a).) While the former case belong to the so-called $p$-wave polar state, the latter is the $f$-wave pairing. That is, dominant triplet pairing symmetry changes as $p$-wave, $f$-wave, $h$-wave,$\cdot\cdot\cdot$, with increasing $p$. To see this more clearly, we conveniently expand the superfluid order parameter as,
\begin{eqnarray}
\Delta(\textbf{p})=\sum_{{\rm odd}~L}\Delta_{L}(p)Y_{L0}\left(\theta_p\right).
\label{Delta_L0}
\end{eqnarray}
Fig.~\ref{fig:SF_L0} shows that each partial wave component $\Delta_L(p)$ take a maximum value at a different $p$, which is larger for a larger angular momentum $L$. We briefly note that this result is quite different form the previous work using $V_{\rm dd}(\textbf{q})$ in Eq.~(\ref{approximate_dipolar_p}), where the $p$-wave polar state is always dominant, irrespective of the value of $p$~\cite{Zhao2010}.
\par
\begin{figure}[t]
\begin{center}\sidecaption
\includegraphics[width=0.4\linewidth]{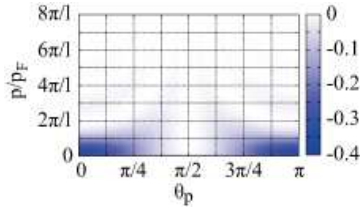}
\caption{(Color online) Calculated self-energy correction $\Sigma(\textbf{p})$ ,as functions of $p$ and $\theta_\textbf{p}$.}
\label{fig:Fock}
  \end{center}
\end{figure}
\begin{figure}[t]
\begin{center}
\includegraphics[width=1.0\linewidth,keepaspectratio]{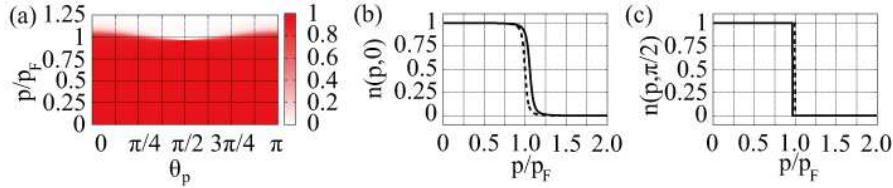}
\caption{(Color online) (a) Calculated momentum distribution $n(\textbf{p})$, as functions of $p$ and $\theta_\textbf{p}$. (b) $n(p,\theta_\textbf{p}=0)$ (polarization direction). (c) $n(p,\theta_\textbf{p}=\pi/2)$ (direction perpendicular to the polarization axis). In panels (b) and (c), dashed lines show the results when $\Sigma\left(\textbf{p}\right)$ is ignored.}
\label{fig:density}
\end{center}
\end{figure}

Figures~\ref{fig:SF}(b) and \ref{fig:SF_L0} also show the results when the self-energy correction $\Sigma(\textbf{p})$ is ignored (dashed lines). These figures show that this mean-field correction enhances $\Delta\left(\textbf{p}\right)$. Because of the anisotropic dipole-dipole interaction, $\Sigma(\textbf{p})$ also becomes anisotropic (See Fig.~\ref
{fig:Fock}.), leading to the deformation of the Fermi surface shape, or the anisotropy of the momentum distribution of dipolar fermions,
\begin{equation}
n(\textbf {p})=\langle a^\dagger_{\textbf{p}}a_{\textbf{p}}\rangle=\frac{1}{2}\left(1-\frac{\xi_\textbf{p}}{E_\textbf{p}}\right),
\end{equation} 
as shown in Fig.~\ref{fig:density}. In particular, panels (b) and (c) indicate that the Fermi surface is elongated along the polarization axis ($p_z$) by $\Sigma(\textbf{p})$, being consistent with the previous works~\cite{Zhao2010,Chan2010,Miyakawa2008}. Then, since the magnitude of $\Delta(\textbf{p})$ is large in the $p_z$-direction, the enhancement of $n(\textbf{p})$ in this direction is favorable to the superfluid phase, which enhances $\Delta(\textbf{p})$, as shown in Figs.~\ref{fig:SF}(b) and \ref{fig:SF_L0}.
\par
%============================================================================
%============================================================================

%--------------------------------------------------------------
\section{Summary}
To summarize, we have discussed the superfluid phase of a polarized dipolar Fermi gas at $T=0$. We have presented an effective dipole-dipole interaction which is free from the ultraviolet divergence, when it is used in the BCS gap equation (without any renormalization). Using this cutoff-free formalism, we clarified the detailed anisotropy of the superfluid order parameter, within the framework of a combined BCS-Leggett theory with the mean-field Fock term correction. When the cutoff-free dipole-dipole interaction is used, the dominant pairing symmetry depends on the magnitude of the momentum $p$, namely, the dominant component changes as $p$-wave, $f$-wave, $h$-wave,$\cdot\cdot\cdot$, with increasing $p$. This result is quite different form previous results using a dipole-dipole interaction involving the ultraviolet divergence, where $p$-wave component is always dominant, irrespective of the magnitude of $p$~\cite{Zhao2010}.
%============================================================================
%============================================================================
\par
Since our approach does not involve any unknown parameter, such as a cutoff energy, when it is extended to a finite temperature including strong pairing fluctuations, quantitative evaluation of the superfluid phase transition temperature $T_{\rm c}$ would be possible, as in the case of superfluid $^{40}$K and $^6$Li Fermi gases~\cite{Nozieres1985,SadeMelo1993,Perali2002,Tsuchiya2009}. Since the prediction of $T_{\rm c}$ is an crucial theoretical issue in the current stage of research for dipolar Fermi gases, our results would be useful toward the accomplishment of this exciting theoretical challenge.

%--------------------------------------------------------------
\begin{acknowledgements}
Y. E. was supported by a Grant-in-Aid for JSPS fellows. Y. O. was supported by Grant-in-Aid for Scientific research from MEXT in Japan (25400418, 25105511, 23500056).
\end{acknowledgements}

%%%%%%%%%%%%%%%%%%%%%%%%%%%%%%%%%%%%%%%%%%%%%%%%%%%%%%%%%%%%%%%%%%%%%%%%%%%%
%\bibliography{DipolarReference}

\end{document}